# A Novel Framework using Elliptic Curve Cryptography for Extremely Secure Transmission in Distributed Privacy Preserving Data Mining

Kiran P[1], S Sathish Kumar[2] and Dr Kavya N P[3]

[1]Department of CS&E, RNS Instituite of Technology, Bangalore
`kiranmys@rediffmail.com`
[2]Department of CS&E, RNS Instituite of Technology, Bangalore
`sathish_tri@yahoo.com`
[3]Prof & Head Department of MCA , RNS Instituite of Technology, Bangalore
`npkavya@yahoo.com`

*Abstract*

*Privacy Preserving Data Mining is a method which ensures privacy of individual information during mining. Most important task involves retrieving information from multiple data bases which is distributed. The data once in the data warehouse can be used by mining algorithms to retrieve confidential information. The proposed framework has two major tasks, secure transmission and privacy of confidential information during mining. Secure transmission is handled by using elliptic curve cryptography and data distortion for privacy preservation ensuring highly secure environment.*

*Keywords*

*Elliptic Curve Cryptograph, Data Distortion and Privacy Preserving Data Mining*

## 1.Introduction

Data mining is used for retrieving intelligent information from huge databases. Presently these databases are distributed across the world.  Distributed data must be retrieved from multiple locations in to the data warehouse, so there is a requirement for a secure transmission and maintaining confidentiality. The transmitted data may contain information which may be private to individual or corporate information which must be secured.

Data mining technology contains tools for automatically and intelligently converting large amount of data in to knowledge relevant to the requirement of an end user. Knowledge discovered by data mining can also retrieve sensitive information about individuals compromising the individual's right to privacy. Moreover, data mining techniques can also reveal important information about business transactions. Thus, there is a sturdy requirement to avoid disclosure of confidential personal information, but also the transmitted data which is considered sensitive in a given context.

The recent research stab is devoted in addressing privacy in data mining. As a outcome of this, a novel group of data mining methods, known as privacy preserving data mining (PPDM), has been developed by the research community. The aim of these algorithms is to retrieve knowledge from data warehouse while preserving confidentiality of data. There are many PPDM methods that have developed in recent years but there is no standardization in these approaches, as discussed in [1]. Privacy preserving data mining typically uses various techniques to modify either the original data or the data generated (calculated, derived) using data mining methods. To achieve optimised results while preserving the privacy of the data subjects efficiently, five aspects or dimensions, have to be taken into account. These dimensions are (1) the distribution of the basic data, (2) how basic data are modified, (3) which mining method is being used, (4) if basic data or rules are to be hidden and (5) which additional methods for privacy preservation are used. This overview shows from a technical perspective how many different methods and techniques in the context of PPDM can be used.

The purpose of this paper is to propose a novel frame work which ensures both secure transmission and confidentiality of information since most of the earlier research has been done on either secure transmission or PPDM method. This frame work can be divided into two major folds. First, we provide a cryptographic method for secure transmission using elliptic curve cryptography. Second, we propose a data distortion method for PPDM mechanism, since we are assuming that we are applying on medical data base which is distributed.

## 2. Literature Survey

Data is distributed across the world, information present in distributed environment must be retrieved to obtain improved data mining result so a secure way of transmission is required. Elliptic curves (EC) were suggested for cryptography by Victor Miller and Neal Koblitz in 1985 as Elliptic Curve Cryptography (ECC) and follows Public Key Encryption technique [2,3]. ECC has a lesser key size, thereby reducing processing overhead. Performance and efficiency of ECC depends mainly on restricted field computations and swift algorithms of elliptic scalar multiplications [4]. Author in [10] presents the implementation of ECC by first transforming the message into an affine point on the elliptic curve (EC), over the finite field GF(p). This paper also illustrates the process of encryption/decryption of text message and also indicates that it is infeasible to attempt a force attack to break the cryptosystem using ECC. In [11] the author has given a multilevel access control for Defence messaging system using Elliptic curve cryptography. The system developed is secure, multi site and allows for global communication using the inherent properties of Elliptic Curve cryptography. It also indicates that Elliptic Curve cryptography provides a greater security with less bit size and it is fast when compared to other schemes. Author in [12] has conducted experiments side-channel attacks and ECC hardware implementations use binary algorithms by observing power consumption of ECC processor on FPGA. Experimental of side-channel attack is conducted to guess the secret key for data encryption and decryption by looking at the physical differences on hardware side effects. In this study, side-channel attack experimental is successful 100% get the key by analyzing of power consumption ECC processor. Author in [13] has indicated a great potential for the future research due to the ever growing demand for compaction of the devices and load balancing for environments in ECC, where there are memory, bandwidth and computational constrains. ECC provides the best way of security for transmission of data [5]. From all the references, ECC has proved to be more secure and uses less key size as compared with all the existing algorithms like RSA.

Privacy preserving data mining is used for secure mining from the data warehouse. The data warehouse can be centralized or distributed. There are various methods for centralized distribution as discussed in [6,18,20]. Random perturbation technique is a method to convert raw data based on probability which has been discussed in [7,8]. Data distortion is achieved by changing the original data, in which some randomness value is added such that the original data is difficult to ascertain, while preserving global feature of a record. In Fixed-data perturbation method the data is changed by adding a noise term e to the attribute X resulting in Y

$$Y = X + e \qquad (4)$$

where e is drawn from some probability distribution [1]. This method is called Additive Data Perturbation(ADP). In Multiplicative Data Perturbation(MDP) the value of e is multiplied with X to get Y the perturbed value,

$$Y = Xe \qquad (5)$$

where e has a mean of 1.0 and a specified variance [9]. In MDP method attribute must be perturbed independently of other attributes. The objective of these data perturbation techniques is to distort the individual data values while preserving the underlying statistical distribution properties. These data perturbation techniques are usually assessed in terms of both their privacy parameters as well as its associated utility measure. While the privacy parameters present the ability of these techniques to hide the original data values, the data utility measures assess whether the dataset keeps the performance of data mining techniques after the data distortion [14]. In this author has investigated the use of truncated non-negative matrix factorization (NMF) with sparseness constraints for data perturbation.

A number of methods have recently been proposed for privacy preserving data mining of multidimensional data records [15]. This paper intends to reiterate several privacy preserving data mining technologies clearly and then proceeds to analyze the merits and shortcomings of these technologies. PPDM has been developed based on Data mining methods. The method of Association Rule Mining in [17] and classification in [16] has been discussed. Author in [21] has discussed Distributed PPDM and also indicated multi-party computation. In all the above mentioned papers either secure transmission or PPDM has been consider. In this paper we are proposing unified approach of ECC for secure transmission and MDP technique for PPDM.

## 3. Proposed framework

The novel framework has been depicted in the fig 1.

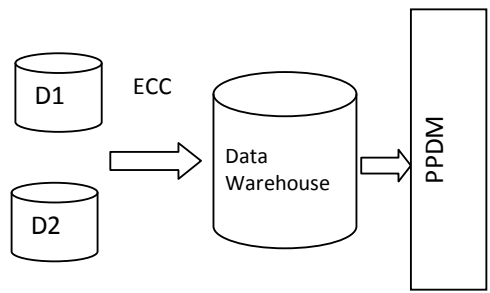

Fig1 Framework embedding ECC and PPDM

Architecture of this unified framework can be divided in to three phases. The first phase is identification of different data sources and encryption of data using ECC before it is transferred on to the Data Warehouse. Next phase is decryption of data which has been send from different data sources to make the data available for Transformation. Transformation is the process of converting data in to appropriate information suitable for Data Warehouse. It also involves cleaning and aggregation of data. Transformed data is loaded on to the next stage. The last Phase contains PPDM approach which uses data distortion to ensure privacy of sensitive information.

## 3.1 ECC for secure transmission

The Elliptic Curve Cryptosystem (ECC), whose security rests on the discrete logarithm problem over the points on the elliptic curve. The main attraction of ECC over RSA and DSA is that it takes full exponential time. RSA and DSA take sub-exponential time. This means that significantly smaller parameters can be used in ECC than in other systems such as RSA and DSA, but with equivalent levels of security.

In realistic terms, the performance of ECC can be increased by selecting particular underlying finite fields of particular interest and referred to as the elliptic group mod p, where p is a prime number. This is defined as follows. Choose two nonnegative integers, u and v, less than p that satisfy:

$$4u^3 + 27v^2 (mod\ p) \neq 0 \qquad (1)$$

Then Ep(x, y) symbolises the elliptic group mod p whose elements (x, y) are nonnegative integers less than p which satisfies the condition:

$$y^2 \equiv x^3 + ax + b(mod\ p) \qquad (2)$$

Together with the point at infinity O. The elliptic curve discrete logarithm problem can be stated as follows. Fix a prime p and an elliptic curve such that

$$Q = xP \qquad (3)$$

Where xP represents the point P on elliptic curve added to itself x number of times. Then the elliptic curve discrete logarithm problem is to determine x given P and Q. It is relatively easy to calculate Q given x and P, but it is very difficult to determine x given Q and P.

D1 and D2 data bases which are distributed across the world. The entire frame can be compared as a three level architecture. Lower level contains the data bases and OLTP. Middle level contains the architecture of data warehouse. The higher level contains PPDM framework. The advantage of this approach is that each level is independent of the other with respect to the process of execution. The initial transfer of information from different sources to the data warehouse is done by means of ECC. The first phase in this system is to encode the data base records as m to be sent as an x-y point $P_m$. The data base D1 and D2 are assumed to be sent by sender S1 and S2. The receiver is R1 who maintains data warehouse. Either S1 or S2 will use the following procedure. It is the point $P_m$ that will be encrypted as a cipher text and subsequently decrypted. Note that we cannot simply encode the message as the x or y coordinate of a point, because not all such coordinates are in $E_p(u, v)$. As with the key exchange system, an encryption/decryption system requires a point G and an elliptic group $E_p(u, v)$ as parameters. Each user S1 or S2 selects a private key $n_A$ and generates a public key

$$P_A = n_A X\ G \qquad (6)$$

To encrypt and send a message $P_m$ to R1, S1 chooses a random positive integer x and produces the cipher text $C_m$ consisting to the pair of points

$$C_m = \{xG, P_m + xP_B\} \qquad (7)$$

Note that S1 has used R1's public key $P_B$. To decrypt the cipher text, R1 multiplies the first point in the pair by R1's secret key and subtracts the result from the second point:

$$P_m + xP_B - n_B(xG) = P_m + x(n_BG) - n_B(xG) = P_m$$
$$(8)$$

S1 has masked the message $P_m$ by adding $xP_B$ to it. Nobody but S1 knows the value of x, so even though $P_B$ is a public key, nobody can remove the mask $xP_B$. However, S1 also includes a "clue," which is enough to remove the mask if one knows the private key $n_B$. For an attacker to recover the message, the attacker would have to compute x given G and xG, which is hard. The same way the data base D2 will be sent by S2 to R1 securely.

Functions of ECC
1. Function to find the multiplicative inverse of an integer for a given prime number:-
For this code, we have used the extended Euclid Algorithm whereby the intermediate terms are less than the prime numbers. This prevents the intermediate terms from exceeding the corresponding prime number.

2. Function to generate the points on an Elliptic curve
As there is constant need for a database of the elliptic curve points, a code to scan all Y co-ordinates that satisfy the elliptic curve equation for the given X co-ordinate has been included.
Equation of the elliptic curve: $y^2 \mod p = (x^3 + ax + b) \mod p$
Where, p is a prime number.
    Algorithm: Inputs: p, a, b
        a. Enter the input data.
        b. x=[0: p-1]
        c. For each value of x, check which values of y from 0 to (p-1) satisfies the equation.
        d. Display the required point.
3. Function to find the public key

4. Function for encoding and decoding

## 3.2 Data distortion for privacy preservation

The D1 and D2 information is securely retrieved from S1 and S2 by R1. The data is transformed to meet the necessary changes of the data warehouse by Extraction Transformation Load (ETL). The second phase of this framework uses data distortion for ensuring privacy during data mining. The MDP algorithm is indicated below

Algorithm
Input: I, N  //The inputs are the vectors of I, composed of confidential numerical attributes only, and the uniform noise vector N, while the output is the transformed vector
subspace I'//
Output: I'

Step 1. For each confidential attribute $A_k$ in       I, where 1<k<d do
    1. Select the noise term $e_k$ in N for    the confidential attribute $A_k$
    2. The k-th operation $op_k \leftarrow \{Mult\}$

Step 2. For each $I_i \in I$ do
    For each $a_k$ in $I_i = I = (a_1,...,a_d)$, where
    $a_k$ is the observation of the k-th attribute
    do
        1. $a'_k \leftarrow$ trasform($a_k, op_k, e_k$)
        Transform($a_k; op_k; e_k$)
End

The result of this data distortion is used for data mining ensuring privacy of the individual data. Association rule is used for mining.

## 4. Experiment

In order to test the performance of our proposed frame work, we conducted a series of experiments on sample medical datasets using matlab as tool. In this section, we present a sample of the results obtained when applying our technique to the original databases using association rule mining.

| No Of Data Records | Accuracy of data mining | |
|---|---|---|
| | Original data set | perturbation data set |
| 200 | 80% | 61% |
| 400 | 85% | 78% |
| 600 | 88% | 82% |
| 800 | 96% | 83% |

Table 1 Comparison with and without PPDM

The experiment result shows that this approach gives a better privacy. This technique is more secure largely because during transmission between distributed data bases to data ware house, ECC is used. Privacy after mining is also retained due to data perturbation. In summary combination of ECC and data perturbation is a good technique for security.

## 5. Conclusion and future work

The growing capacity to track and collect huge amounts of data which is distributed across different locations has lead to an interest in the development of reliable proper security over communication and also data mining algorithms which preserve user privacy. In this paper, we have proposed a novel framework for privacy preserving data mining for distributed database combining ECC and data distortion technique. ECC has been used for secure transmission of data from source to destination which ensures security during transmission and data distortion mechanism for PPDM. This unified approach gives better security and privacy of data as compared with rest of the approaches in literature as most of the earlier research was either on to secure transmission or on to PPDM approach.

Many motivating and imperative directions are worth exploring. This approach can also be used as a framework for the next generation researchers on Data Warehouse since most of the present Data Warehouse are distributed which requires security during transmission and privacy of confidential information. We have used data distortion mechanism for PPDM but there are also other approaches in this area, such as random rotation based data perturbation, *k*-anonymity, and retention replacement methods, it is interesting to analyze how these methods can be used in the above frame work. It is also of great interest to extend our approach to handle evolving data streams.

Future research also includes taking actual data and analysing communication speed and the overhead involved by adding ECC on PPDM. In our framework, Association rule mining is used there is a need to have PPDM mechanism to be designed for other mining algorithms like clustering also. Further research may focus on the combination of randomization and cryptographic methods to get better results in PPDM ensuring high Data Mining results.